\documentclass[a4paper,fleqn,usenatbib]{mnras}

\usepackage{newtxtext,newtxmath}

\usepackage[T1]{fontenc}
\usepackage{ae,aecompl}
\usepackage{pdflscape}

\usepackage{graphicx}	
\usepackage{amsmath}	
\usepackage{amssymb}	

\title[Evolving RCB Stars with {\it MESA}]{Evolving R Coronae Borealis Stars with {\it MESA}}

\author[A. Lauer et al.]{Amber Lauer$^{1,2}$,
Emmanouil Chatzopoulos$^{1}$,
Geoffrey C. Clayton$^{1}$\thanks{E--mail: gclayton@fenway.phys.lsu.edu}, 
\newauthor Juhan Frank$^{1}$,
and Dominic C. Marcello$^{1}$
\\
$^{1}$Dept.\ of Physics \& Astronomy, Louisiana State University, Baton Rouge, LA 70803, USA\\
$^{2}$Triangle Universities Nuclear Lab, Duke University, Durham, NC 27710, USA
}

\date{Accepted XXX. Received YYY; in original form ZZZ}

\pubyear{2018}

\begin{document}
\label{firstpage}
\pagerange{\pageref{firstpage}--\pageref{lastpage}}
\maketitle

\begin{abstract}
The R Coronae Borealis (RCB) stars are rare hydrogen--deficient, carbon--rich supergiants. 
They undergo extreme, irregular declines in brightness of many magnitudes due to the formation of thick clouds of carbon dust. It is thought that RCB stars result from the mergers of CO/He white dwarf (WD) binaries.
We constructed post--merger  spherically--symmetric models 
computed with the MESA code, and then followed the evolution into the region of the HR diagram where the RCB stars are located. We also investigated nucleosynthesis in the dynamically accreting material of CO/He WD mergers which may provide a suitable environment for significant production of $^{18}$O and the very low $^{16}$O/$^{18}$O values observed.
We have also discovered that the N abundance depends sensitively on the peak temperature in the He--burning shell.
Our {\it MESA} modeling consists of engineering the star by adding He--WD material to an initial CO--WD model, and then following the post--merger evolution using a nuclear--reaction network to match the observed RCB abundances as it expands and cools to become an RCB star.  
These new models are more physical because they include rotation, mixing, mass-loss, and nucleosynthesis within {\it MESA}.
We follow the later evolution beyond the RCB phase to determine the stars' likely lifetimes. The relative numbers of known RCB and Extreme Helium (EHe) stars correspond well to the lifetimes predicted from the {\it MESA} models. In addition, most of computed abundances agree very well with the observed range of abundances for the RCB class.

\end{abstract}




\section{Introduction}

The R Coronae Borealis (RCB) stars make up a class of rare hydrogen--deficient supergiants that may be the result of a white--dwarf (WD) merger event \citep{2012JAVSO..40..539C}.
R CrB is the first member of the RCB class having been discovered as variable over 200 yr ago \citep{1797RSPT...87..133P}.
The RCB stars are best known for sudden declines in brightness of 8 magnitudes or more at irregular intervals caused by clouds of carbon dust forming near the atmosphere of the stars, which are later dissipated by radiation pressure. Only about 120 RCB stars are known in the Galaxy \citep{2012JAVSO..40..539C,2013A&A...551A..77T,2018arXiv180901743T,Shields_2019}.  There are also five hydrogen--deficient carbon (HdC) stars that mimic RCB stars spectroscopically, but do not show declines in brightness or IR excesses \citep{Warner:1967lr}. The RCB stars may be the result of a rare form of stellar evolution or exist in an evolutionary phase that lasts only a short time. In addition to their extreme hydrogen deficiency, RCB stars show uniquely low $^{16}$O/$^{18}$O, large $^{12}$C/$^{13}$C, and enhanced s--process elements which are all consistent with partial He-burning \citep{Clayton:2007ve}. 
Two scenarios have been suggested for producing an RCB star: the double degenerate and the final helium--shell flash models \citep{Iben:1996fj,2002MNRAS.333..121S}. In the former, an RCB star is the result of a merger between a CO-- and a He--white dwarf (WD)
\citep{1984ApJ...277..355W}. In the latter, a star evolving into a WD undergoes a final helium--shell flash and expands to supergiant size \citep{Fujimoto:1977lr}. 
The preponderance of the evidence seems to support the WD--merger scenario \citep{Clayton:2007ve,2011ApJ...743...44C,2012JAVSO..40..539C}. 



We use the 1D {\it MESA} (Modules for Experiments in Stellar Astrophysics) stellar evolution code to construct spherical models after the merger of a CO-- and a He--WD \citep{staff2012, 2013ApJ...772...59M,staff2018}. With {\it MESA}, we follow the evolution into the region of the HR diagram where RCB stars are located. 
We present our study of the evolution of 0.8--1.05 $M_{\odot}$ post--merger objects with the aim of reproducing the observed nuclear abundances of RCB stars and several characteristic timescales \citep{staff2012, 2013ApJ...772...59M,staff2018}. Our {\it MESA} modeling consists of two steps. First, we mimic the WD merger event by applying ``stellar engineering" to an initial CO--WD model by wrapping it in He--WD material and then applying an entropy adjusting procedure \citep{2012ApJ...748...35S,Schwab_2016} to expand the He--rich envelope to a radius consistent with that found in past 3D hydrodynamics simulations. 

Second, we follow the post--merger evolution using a 75--isotope nuclear--reaction network that is co--processed within the evolutionary calculations, including the effects of convective and rotational instabilities to the mixing of material, in an attempt to match the observed RCB abundances. {\it MESA} follows the evolution of the merger product as it expands and cools to become an RCB star. We then examine the surface abundances and compare them to the observed range of RCB abundances \citep{Asplund:2000qy,Clayton:2007ve,Hema_2017}.
We also investigate how long fusion continues in the He shell near the core and how this processed material is mixed up to the surface of the star, as well as the timescale for a post--RCB model to reach the Extreme He star (EHe) phase, and the timescale to cool and return to the WD phase. We then use these timescales to estimate the RCB population.

This work is an improvement on previous studies using grid and SPH hydrodynamics codes as well as {\it MESA} \citep{2011ApJ...737L..34L,staff2012,2013ApJ...772...59M,2014MNRAS.445..660Z,staff2018,Menon_2018}. We compare our results to these previous studies and emphasize the advantages of our approach which include rotation, mixing, mass-loss during the pre-RCB and RCB phases, and
nucleosynthesis within MESA.


\setcounter{table}{0}
\begin{table}
\caption{``Large'' ({\tt mesa\_75.net}) Isotope Network}
\begin{tabular}{llllll}
\hline
Element&
A$_{min}$&
A$_{max}$&
Element&
A$_{min}$&
A$_{max}$ \\ \hline
n&1&1&&&\\
H&1&2&S&31&32\\
He&3&4&Cl&35&35\\
Li&7&7&Ar&35&38\\
Be&7&10&K&39&39\\
B&8&8&Ca&39&42\\
C&12&13&Sc&43&43\\
N&13&15&Ti&44&46\\
O&14&18&V&47&47\\
F&17&19&Cr&48&50\\
Ne&18&22&Mn&51&51\\
Na&21&24&Fe&52&56\\
Mg&23&26&Co&55&56\\
Al&25&27&Ni&56&58\\
Si&27&28&Cu&59&59\\
P&30&31&Zn&60&60\\
\hline
\end{tabular}
\label{T1}
\end{table}

\setcounter{table}{1}
\begin{table}
\label{net_table}
\caption{``Reduced" ({\tt sagb\_NeNa\_MgAl.net}) Isotope Network}
\begin{tabular}{lll}
\hline
Element&
A$_{min}$&
A$_{max}$
\\
\hline
n&1&1\\
H&1&2\\
He&3&4\\
Li&7&7\\
Be&7&8\\
B&8&8\\
C&12&13\\
N&13&15\\
O&16&18\\
F&19&19\\
Ne&20&22\\
Na&21&23\\
Mg&24&26\\
Al&25&27\\
\hline
\end{tabular}
\label{T2}
\end{table}

\begin{figure*}
\begin{center}
\includegraphics[angle=0,width=4in]{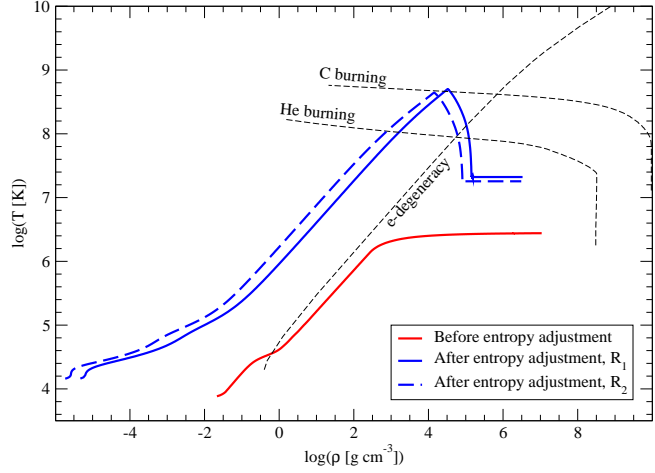}
\caption{Density--temperature ($\rho$--$T$) structure diagram of the He/CO--WD post--merger object before 
(red curve) and after entropy adjustment for two cases: a case leading to a smaller radius ($R_{\rm 1}$; solid blue curve)
where the peak temperature of the hot shell exceeds the lower limit for carbon burning and a case leading to a larger
radius ($R_{\rm 2}$; dashed blue curve) where the hot shell is entirely within the helium--burning regime. The dashed black
curves indicate the $\rho$--$T$ limits for electron degeneracy, He and C burning.}
\label{Fig:MergerEngineering}
\end{center}
\end{figure*}

\begin{figure*}
\begin{center}
\includegraphics[angle=0,width=4in]{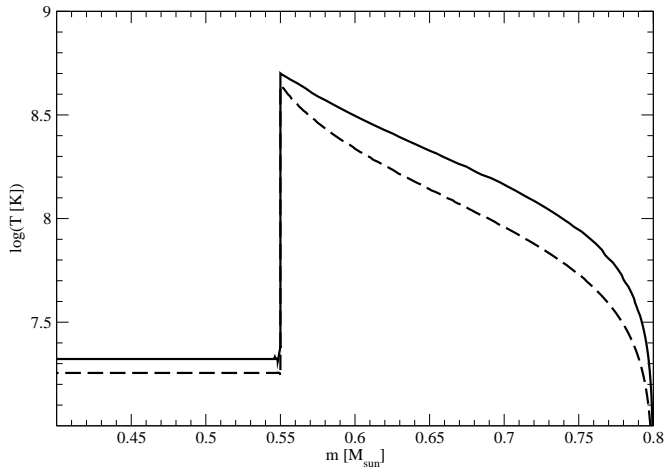}
\caption{Temperature profiles for the final, entropy--adjusted post--merger objects illustrating the He--burning shell defining
the interface between the CO and the He--WD material. Two cases are shown, one for a compact (radius $R_{\rm 1}$; solid curve)
and one for a more extended (radius $R_{\rm 2}$; dashed curve) post--merger envelope with a cooler He-burning shell.}
\label{Fig:TemperatureProfile}
\end{center}
\end{figure*}

\section{Constructing CO/He white dwarf post--merger Objects with {\it MESA}}\label{StellEng}

To simulate the evolution of RCB stars, we use the {\it MESA} 1D, radial, spherically--symmetric, stellar--evolution code \citep{Paxton_2010,Paxton_2013,Paxton_2015}. Therefore, there are some concessions in terms of reproducing a realistic structure of the CO/He--WD post--merger object. As shown by 3D hydrodynamic simulations,
the structure of such an object right after merger is composed of three components which demonstrate radial and/or angular asymmetry:  a cold core (the accreting CO--WD), a hot, puffy corona, and a disk (composed of the disrupted He--WD material) \citep{staff2012,staff2018}. The photosphere will not have a spherical but rather an elongated shape, as shown in multiple hydrodynamics simulations \citep{2002MNRAS.333..121S,2007ApJ...670.1314M,2007MNRAS.380..933Y,2011ApJ...737L..34L,staff2012,2013ApJ...767..164Z,2014MNRAS.445..660Z}.

However, as described in Section 5, the post--merger object becomes spherical in only a few years. Assuming that the post--merger object has reached a spherical shape, we apply the ``stellar engineering'' methods discussed below in order to: (a) constrain the initial abundance profiles for both the CO-- and the He--WD before the merger based on two different nuclear reaction networks (a 75--isotope network and a 29--isotope network) and (b) adjust the structure of the post--merger object that will be used as the initial configuration for the evolutionary calculations. 

\subsection{Computing the initial abundance profiles of the progenitor WDs}\label{StellEng1}

To produce the initial abundance profiles for the He-- and CO-- WD models, we 
used the provided {\it MESA} test--suite cases, {\it make\_he\_wd} and {\it make\_co\_wd}, coupled with our choice of two nuclear reaction networks ({\tt mesa\_75.net and sagb\_NeNa\_MgAl.net} corresponding to a 75--isotope and a 29--isotope network respectively).  

 In order to compute the CO--WD model, {\it MESA} evolves a 3.25~$M_{\odot}$, solar metallicity star using a CNO network up to the first thermal pulse, and then changes the wind prescription until the result is a 0.6~$M_{\odot}$ CO WD. For the He--WD, a model with initial mass of 1.5~$M_{\odot}$ model is evolved through the main sequence and then relaxed into a 0.15 ~$M_{\odot}$ model (while keeping the compositional profile intact) in the {\tt make\_he\_wd} test problem supplied in {\it MESA}. This is applied via the ``relax\_mass" method available in {\it MESA}. It is then the subsequent evolution of the 0.15~$M_{\odot}$ model that yields the He WD. For this test problem, an artificial mass--loss is applied to get to the desired small mass. In both test--suite cases, the WD stage that is adopted corresponds to the phase when the luminosity drops to $L=10^{-2}$~$L_{\odot}$.
 

Following this procedure, the CO-- and He--WD compositional structures are computed independently in {\it MESA} for each network choice: the ``large'' ({\tt mesa\_75.net}) network with 75 species ranging from neutrons to $^{60}$Zn (Table~\ref{T1}) and the ``reduced'' ({\tt sagb\_NeNa\_MgAl.net}) network with 29 species ranging from neutrons to $^{27}$Al (Table~\ref{T2}). The ``reduced'' network is chosen because it is a subset of the ``large'' network and allows for faster computing times. The final CO--WD is computed using a radial mass fraction profile. For the He--WD, we compute abundance values averaged over the entire mass of the model star assuming that the disrupted He--WD material will be heavily mixed in the process of becoming the corona of the post--merger object. This results in a uniform He WD abundance profile (i.e. the species mass fractions are constant for all mass coordinates).


\subsection{Engineering a CO+He WD post--merger structure}\label{StellEng2}

To create an approximate CO/He--WD post--merger structure and set the initial conditions for the {\it MESA} evolution calculation, we follow a multi--step process similar to that used by \citet{2012ApJ...748...35S} and \citet{Schwab_2016}. First we numerically relax a model with mass equal to the desired mass of the post--merger object and with nuclear burning and mixing turned off until the core (corresponding to the compact CO-WD part of the post--merger object) becomes degenerate (see Table~\ref{T3}).
We designate that phase to be when $\eta_{\rm c} =$~0.5, where $\eta_{\rm c} \sim \mu/k_{\rm B} T$ and $\eta_{\rm c}$ is the degeneracy parameter at the boundary of the central core, $\mu$ the electron chemical potential and $T$ the temperature. Second, we restart the calculation in order to adjust the composition of the entire core to one corresponding to the CO--WD abundance calculated independently in {\it MESA} during the steps described in Section 2.1. This ensures that the composition of the CO--WD will be the same as one computed by running the {\it make\_co\_wd MESA} test--suite problem by using either the ``large'' or the ``reduced'' network option. The third step is to allow the now degenerate model to cool down to a core temperature of $\sim$1.6 $\times 10^{7}$~K. In the fourth step, we select the mass coordinate that sets the desired boundary between the CO-- and the He--WD portions of the post--merger object (set by the individual masses of the CO--WD and the He--WD) and adjust the composition from that point out to the surface using the pre--computed homogeneous composition profile for the He--WD.  We note that in this case, the abundance profile of the He WD component is uniform (since it corresponds to the mass--averaged (mixed) profile of the He WD computed in Section~\ref{StellEng1}.

At this stage, we have a composite fully degenerate CO/He object. Thus, the final step is to adjust the structure to mimic the size and characteristics of the cold CO core and puffy He corona that make up the post--merger object following the fast and slow accretion phase seen in hydrodynamic simulations of CO/He--WD mergers \citep{staff2012,2014MNRAS.445..660Z}.
To do so, we use an entropy adjusting procedure as detailed in \citet{2012ApJ...748...35S}. Entropy is added to the envelope until it expands to a desired radius while nuclear burning, neutrino cooling, and chemical mixing are kept turned off.
The physical limits for the final size (radius) and rotational velocity of the post--merger objects are based on 3D hydrodynamical merger simulations \citep{staff2012,staff2018}. In general we adopt a post--merger object radius of about 0.1 $R_{\odot}$ and an initial equatorial rotational velocity of 20\% the critical Keplerian value corresponding to break--up velocity, values that are consistent with 3D hydrodynamics simulations done in the past. 

Figure~\ref{Fig:MergerEngineering} shows the engineered $T$--$\rho$ structure of the 0.8~$M_{\odot}$ He/CO--WD post--merger object before and after the envelope entropy injection phase. Figure~\ref{Fig:TemperatureProfile} shows the temperature ($T$) profile of the final post--merger object for two cases (low and high entropy injection). This Figure can be directly compared to Figure 4 of \citet{2014MNRAS.445..660Z}. The amount of entropy injected into the envelope sets the initial radius as well as peak temperature in the hot shell of the post--merger object. Following the entropy adjustment process and depending on how much entropy is injected into the envelope of the post--merger object, the radius of the He--WD corona will increase accordingly. In Figures~\ref{Fig:MergerEngineering} and \ref{Fig:TemperatureProfile}, two examples are shown, the ``R2" case resulting from 10\% more entropy injected than in the ``R1" case. Higher entropy and a more extended envelope result in a somewhat lower peak temperature (Figure~\ref{Fig:TemperatureProfile}) and less C--burning (Figure~\ref{Fig:MergerEngineering}) in the hot, He--burning shell. Note how the tip of the density--temperature profile barely edges into the C--burning region for the ``R2" case.


\section{The Nuclear Network}
One of the primary improvements of this model over previous works is the incorporation of multi-zone, co-processing of the nuclear network. Most previous models of RCB stars, and indeed other stellar models, relied on post--processing methods \citep{2011ApJ...737L..34L,2012A&A...542A.117L,2013ApJ...772...59M,Menon_2018}, which treat the nuclear network separately from the stellar evolution.  \citet{2014MNRAS.445..660Z} also employed nucleosynthesis within {\it MESA} but using a small network which didn't include neutron reactions. Table~\ref{T99} compares the models used by these studies.
The Longland and Menon studies relied on post-processing. In post--processing methods, a truncated nuclear network is used in the stellar evolution. If an extensive nucleosynthesis profile is desired, it is modeled separately in dedicated software including hundreds or thousands of species. A thermodynamic profile evolved in time is derived from the stellar evolution (SE) software and applied to one or more abundance zones. 
The reduced nuclear networks used in the stellar evolution phase of post-processing contain just a few species beyond hydrogen. Many sources demonstrate evidence that this accounts for most of both the nucleosynthesis and energy generation \citep[]{Paxton_2013}. However, small networks used in this manner often ignore species with low mass fractions and reactions with low cross sections.  
Nuclear reactions are the primary driver of the thermodynamics of stars, while their cross sections are dependent on these thermodynamics. The interplay between the two is a vital part of the model. Thus the feedback between the full network and the stellar evolution, especially in exotic environments, is important and the results could deviate significantly from those of a truncated network. In addition, it is important to account for the impact of rotation and mixing, which are discussed in other sections, and their effects on the nucleosynthesis, energy, and abundance (also see \citet{2013A&A...559A...4C} and references therein.


Recently {\it MESA} introduced the capability to include nuclear reaction networks {\it in situ} that are fully--coupled to the stellar evolution (rather than using the uncoupled co--processing approach). In this approach, the system of coupled equations of stellar evolution is expanded to include the full nuclear network equations, which are then concurrently evolved (see \citet{2015ApJS..220...15P} and relevant discussion in their section 5.2). In other words, the complete system of equations for all structure variables and all chemical species is solved at the same time. Also, as {\it MESA} calculates stellar evolution in multiple, often thousands of zones, and the mass fractions and network are calculated in each of these with mixing between. This is the reason we opted to rely on the results of the {\it MESA} network computations rather than an additional post-processing network. The computational cost of this is significant, however. As such, it is important to choose the correct network within {\it MESA}. The primary goals of this study are to match the RCB observations \citep{Asplund:2000qy,2011MNRAS.414.3599J}, and also to compare our new results with previous {\it MESA} models \citep{2011ApJ...737L..34L,2013ApJ...772...59M,2014MNRAS.445..660Z}. 
The increased capabilities of {\it MESA} make it possible to undertake such an exercise using a single computational package.


Of special interest in RCB class stars are $^{12}$C/$^{13}$C, $^{16}$O/$^{18}$O, F, Li, and various s--process elements. RCB elements important in the s--process include H, Li, C, N, O, Na, Al, Si, S, Ca, Sc, Fe, Ni, Zn, Y, and Ba. The optimal network would include all of these species as well as the intermediate nuclei, as the majority of reactions on nuclei involve fusion of a single particle or alpha particle. However, in this scenario completeness competed with time constraints to determine the final network choice. A network including relevant stable species of Ba, as well as the intermediate isotopes, would contain nearly 2000 species and 20,000 reactions, which is too large for our computing resources. Although {\it MESA} is parallelizable via openMP, it is limited to one machine. A test run was done to benchmark a large network using the A1 model. The net included ~800 species up to Y. After several hundred hours on a x16 Xeon processor, and 46 Gb of RAM, the model star had only evolved ~10$^{-4}$s. The final network chosen was {\tt mesa\_75.net}. This is includes relevant species up to $^{60}$Zn (see Table \ref{T2}). The goal was to include as many species as possible while still preserving computational speed. Note that the networks used here do not include the abundant, stable isotopes for all the elements in Tables~\ref{T1} and ~\ref{T2}. In particular, the abundances of $^{11}$B, $^{45}$Sc, $^{51}$V, $^{55}$Mn, $^{59}$Co, $^{63}$Cu, and $^{64}$Zn are not estimated by our models. While it is always preferable to include isotopes whenever possible, it should be noted that these species are known to be created only by exotic processes not expected to occur in this environment. 
The chosen network is complete in that the nuclear energy production, the reaction rates and, as a result, the computed isotopic abundances for the species included are suitable for the helium--burning regime that is relevant to this work in the context of RCB stars and for modeling the abundances that we are comparing against their observations. 


\setcounter{table}{2}
\begin{table*}
\caption{Study Comparison}
\begin{tabular}{llllllcc}
\hline
Study&
Nucleosynthesis&
Network&
Neutrons&
Li&
Rotation&
Mixing&
Mass loss\\
&&&&&&&pre-RCB/RCB\\
\hline
Longland et al. (2011, 2012)&SPH (PP)&327&yes&yes&no&yes&no/no\\
Menon et al. (2013, 2018)&{\it Nugrid} (PP)&$\sim$1000&yes&no&no&adhoc&no/yes\\
Zhang et al. (2014)&{\it MESA}&35&no&yes&no&MLT&no/yes\\
This study&{\it MESA}&75&yes&yes&yes&MLT&yes/yes\\
\hline
\end{tabular}

\label{T99}
\end{table*}

\section{Comparison with Previous Work}

In this section, we discuss the key differences and improvements between our method and previous theoretical studies of RCB stars and, in particular, with the works of \citet{2011ApJ...737L..34L,2012A&A...542A.117L}, \citet{2013ApJ...772...59M,Menon_2018} and \citet{2014MNRAS.445..660Z}. The differences, summarized in Table~\ref{T99}, span all stages of constructing and evolving a He/CO--WD post--merger object, from the initial structure, spatial resolution and dimensionality of the post--merger object, to the nuclear reaction network size and nuclear energy generation, chemical mixing and mass--loss.

\subsection{Initial structure of the He/CO--WD post--merger object.}

When setting the initial conditions for the evolution of a He/CO--WD post--merger object it is important to consider the dynamics of the merger itself and the realistic, three--dimensional properties of the resulting structure. As described in detail in Section 2, we approximate this step in 1D spherical symmetry with {\it MESA} by adjusting the entropy of the He--rich corona, mimicking post--merger structures computed in 3D hydrodynamics \citep{staff2012, staff2018}. The advantage of this method is that it is computationally economical and allows for good resolution of the entire post--merger object (our initial post--merger structures have several 1000s of zones). 

In contrast, \citet{2011ApJ...737L..34L} adopted previously published 3D SPH simulations of 0.8+0.4~$M_{\odot}$ CO/He--WD mergers leading to RCB stars. These resulted from fast and slow accretion phases producing a core (CO--WD) surrounded by a He corona and an on--going accretion disk. Over the course of the 3D SPH merger simulations, thermodynamic profiles of tracer ``particles" consisting of stellar material are tracked and individually used in the post-processing calculations. The final abundances are calculated from the tracer particles still residing in the photosphere in the final model.

\citet{2013ApJ...772...59M,Menon_2018} produced the first 1D {\it MESA} model of RCB stars although their post--merger object abundance profile was somewhat contrived.
They used a four--zone model that consisted of a core based on the pre--merger CO--WD, a shell of mixed He--WD material, and two additional zones between them: the so--called shell--of--fire where temperatures are high enough for He--burning \citep{staff2012}, and a non--physical ``buffer'' zone. The buffer zone is 97\% He with trace amounts of H and N, while the shell--of--fire consists of a mixture of material from the He-- and CO--WDs.

\citet{2014MNRAS.445..660Z} used {\it MESA} to evolve a 4~$M_{\odot}$ (solar metallicity) model on the ZAMS until the H core had a mass of .65~$M_{\odot}$, and then gradually stripped the outer layers (with the {\it MESA} {\tt relax\_mass} option). This bare H core was evolved to He with a mass of 0.61~$M_{\odot}$. The final CO--WD was made by removing the rest of the envelope around the core and allowing evolution to continue while cooling, until the lower limit of the luminosity reached 10$^{-2}$~$L_{\odot}$. 
The merger was mimicked by a few computational steps with slow accretion, ~(10$^{-4}$~$M_{\odot}$ yr$^{-1}$) followed by a fast (10$^{4}$~$M_{\odot}$ yr$^{-1}$) accretion phase in which He--WD abundance material joins the 0.6~$M_{\odot}$ CO--WD remnant. 

The fast accretion used by \citet{2014MNRAS.445..660Z} is no longer numerically stable in {\it MESA} due to the fact that it was unphysical, and improvements in the accompanying code now limit maximum accretion to
well below the theorized rates of the merger.

All of the studies including ours employ solar abundances as a starting point. \citet{2011ApJ...737L..34L} and \citet{Menon_2018} also ran models with sub-Solar abundances.


\subsection{Nucleosynthesis}
In contrast with these previous studies, the post--merger abundance profile used in this work was derived by following the evolution of the progenitor stars with the 75--isotope and the 29--isotope co-processed networks used from the very first step of the {\it MESA} runs (Section 2.1), including rotation and mixing post-merger in subsequent evolution phases.

In their work, \citet{2013ApJ...772...59M} used {\it MESA} version 3851 with the built-in network agb.net, which includes 19 species and 27 reactions. This resulted in a thermodynamic profile of 1300 physical zones, which was input into a {\it NuGrid} nucleosynthesis calculation of over 1000 isotopes and including mixing\footnote{http://www.nugridstars.org}. In the {\it MESA} portion of the work, the authors used an abundance profile which did not include all isotopes of the network. The primary concern with varying the networks from model to model is that species inconsistent between the starting abundances and the network list are reassigned in a way that may not reflect their actual values \citep{Paxton_2015}. 
\citet{Menon_2018} use {\it MESA} version 6794 but otherwise their method is unchanged. However, they do include additional models that treat the hot-merger scenario, or what they refer to as the Shell-of-Fire. 

\citet{2014MNRAS.445..660Z}  used {\it MESA} version 4028, with a custom network, containing 35 species and 113 reactions,  including the following: H(1--2), He(3--4), Li(7), Be(7), B(8), C(12--13), N(13--15), O(14--18), F(17--19), Ne(18--22), Mg(22--24), Na(23), Al(27), P(31), and S(32), but didn't include neutrons. The is an important difference as discussed below in Section 5.2.4. They used a {\it MESA} hard--coded reaction rate for 3$\alpha$~reactions based on \citet{Fushiki_1987}, and the $^{12}$C+$^{12}$C reaction rate based on \citet{Gasques_2005}. The same standard {\it MESA} rates are used here. 
 
In two separate studies, \citet{2011ApJ...737L..34L, 2012A&A...542A.117L} relied on independent SPH models, calculated  previously, to obtain the thermodynamic input for post-processing calculations. The stellar evolution included a 14 species network which assumed the nucleosynthesis was dominated by alpha-capture which is thought to dominate the energy generation. \citet{2011ApJ...737L..34L} suggest this energy generation only amounts to a 1\% difference from the larger network calculated subsequently. The post-processing calculations utilized 327 species up to Ga 70, in a hybrid multi-zone approach adapted for SPH. Tracer ``particles", which are regions of stellar material of mass $\sim10^{-6}$~$M_{\odot}$ in the star were treated individually with nuclear networks. The results were culled to include only those that were located in the region outside the hot corona. The remaining particles were then summed to find the final reported mass fraction. Although the dynamical merger simulations of \citet{2011ApJ...737L..34L} were done in 3D, and the nuclear post-processing calculations followed thousands of tracer ``particles" representing large stellar volumes, these particles did not interact with each other during nuclear evolution, and so do not include coupled effects of mixing and rotation.

The models presented in our study are more consistent throughout the process, utilizing a single stellar evolution code, which includes a co-processed network, as well as the effects of mixing and rotation. We also follow the evolution from the progenitor stars, to post-merger, to RCB star, to WD with one streamlined process in which we mimic realistic physical conditions as much as possible. Though the nuclear net used in this work contains fewer isotopes (75), all of those isotopes ultimately provide feedback between cross section and thermodynamics throughout the entirety of the model. The new models presented here have self--consistent evolution of the post--merger object and nucleosynthesis with energy generation for 75 isotopes and other important properties (mixing, rotation, mass--loss). We also predict time--scales and stellar evolution tracks. As we show later, our models naturally get to the ``correct" (i.e., in agreement with observations) mass when they reach the RCB phase. 


\subsection{Chemical mixing.}

The treatment of mixing during the evolutionary calculation of the post--merger object is another important aspect of physics that needs to be considered since empirical evidence suggest that the surface abundances of RCB stars are rich in products of He--burning that need to be transferred from the inner and hotter, He--burning shell to the surface of the star. In our study, we are making use of the provided {\it MESA} mixing prescriptions, and more importantly, convective mixing as formulated in the 1D Mixing Length Theory (MLT). We adopt $\alpha_{\rm MLT}=$~2.0 which is identical to the choice of \citet{2014MNRAS.445..660Z} who also used the provided {\it MESA} mixing routines. In contrast, \citet{2013ApJ...772...59M,Menon_2018} implement a custom, non-physical mixing routine in {\it MESA} to enhance mixing efficiency in their four--zone model.
To quantitatively follow convective mixing in {\it MESA}, the convective velocity that is obtained from the MLT equations, $v_{\rm conv}$, is used to calculate a convective diffusion coefficient ($D_{\rm conv}$) that is then added to the diffusion equations that the code uses to calculate mixing by different processes \citep{2011ApJS..192....3P}. As such, mixing is time--dependent in our models.

\citet{2011ApJ...737L..34L} on the other hand, calculate the mass--averaged abundances of all particles that reside within a defined convective zone at the end of the hydrodynamic simulation and leave the depth of the mixing as a free parameter in order to investigate the possibility that the entire hot He--rich corona is convective. They consider two cases: a case of ``deep'' mixing (where everything in the radial coordinate range 0.005~$< R/R_{\odot} <$~0.05 is mixed homogeneously) and a case of ``shallow'' mixing (where the homogeneous mixing range is smaller; 0.014~$< R/R_{\odot} <$~0.05).

In our treatment, we also include the effects of rotationally--induced mixing since we assign an initial equatorial velocity equal to 20\% of the critical Keplerian value, a value that is consistent with the one suggested by 3D hydrodynamic merger simulations \citep{staff2012}. During the RCB phase, rotationally--induced mixing is dominated by meridional circulation that is active between the hot He--burning shell and the extended He--corona. We show in Section~5.2.4 that mixing due to rotation does have a limited yet robust effect on the computed RCB surface abundances. 


\subsection{Mass--loss.}

Wind--driven mass--loss during the evolution of the He/CO--WD post--merger object is another important parameter and it is very uncertain in 1D stellar evolution. {\it MESA} allows the user to choose from a variety of available mass--loss prescriptions that span different regimes of effective temperature and metallicity. In our work, we adopt the  Reimers and Bl\"{o}cker mass--loss method throughout the evolution calculation which is the same used by \citet{2014MNRAS.445..660Z}. A key difference however is that \citet{2014MNRAS.445..660Z} do not turn on mass--loss until their models reach the RCB phase in order to get final RCB masses in agreement with observations (an example is shown in Figure 7 of \citealt{2014MNRAS.445..660Z}). In our work we keep mass--loss turned on throughout the entire evolution from the He/CO--WD post--merger initial structure to the RCB phase albeit with a smaller efficiency for the Bl\"{o}cker part. As shown in Tables~\ref{T3}--\ref{T4}, the stellar mass remains high ($\sim$0.8~$M_{\odot}$ for the A--models) at the start of the RCB phase and most of the mass-loss occurs during the RCB phase.

The Bl\"{o}cker wind--driven mass--loss formula is also adopted by \citet{2013ApJ...772...59M} with an efficiency set to $\eta =$~0.05, similar to what is used in their {\it NuGrid} Asymptotic Giant Branch (AGB) simulations. In comparison, \citet{2011ApJ...737L..34L} do not consider any mass--loss during the nucleosynthetic phase and only allow for the small mechanically--driven mass--loss associated with the merger process.

\begin{figure*}
\begin{center}
\includegraphics[angle=0,width=6in,clip]{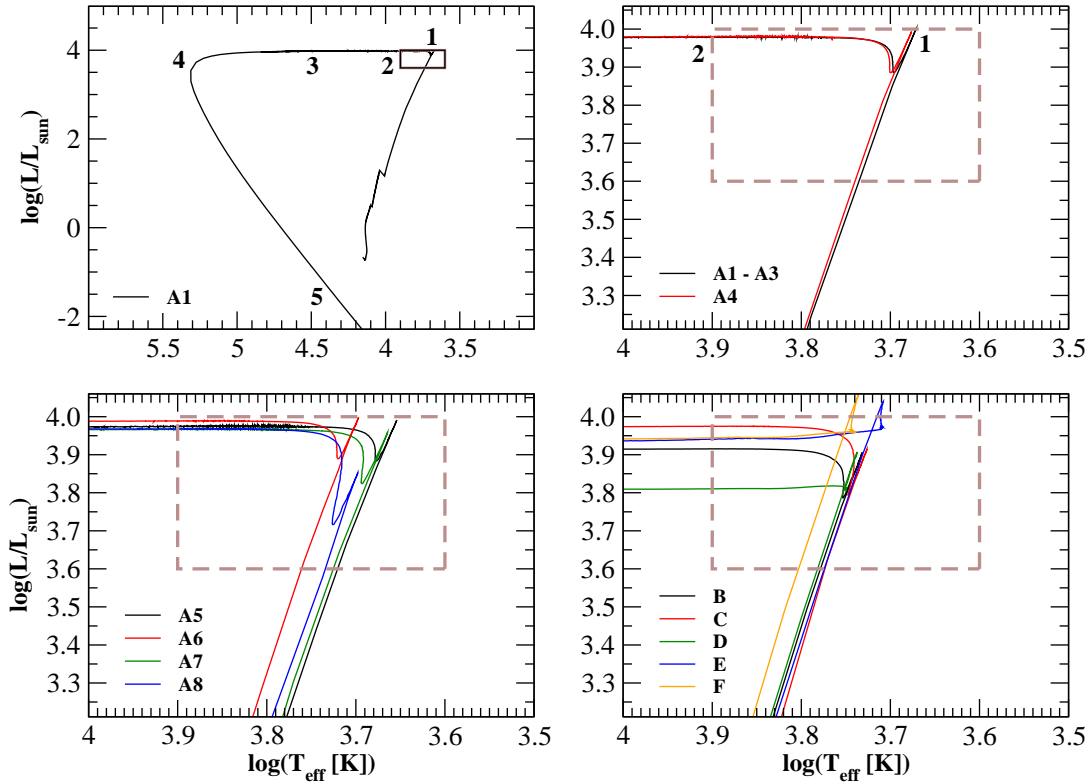}
\caption{{\it Top left panel}: Evolution in the H--R diagram of the 0.8~$M_{\odot}$ CO/He--WD post--merger object for model A1. The brown box indicates the observed range of L and $T_{\rm eff}$ for RCB stars. The number labels correspond to the following evolutionary phases: (1) RCB phase (taken to be when the model reaches its maximum luminosity), (2) End of the RCB phase (when $\log T_{\rm eff} =$~3.9), (3) EHe stars ($\log T_{\rm eff} =$~4.5), (4) the end of the nearly--constant luminosity post--RCB phase (when $\log L$ starts to decrease) and (5) WD phase (taken to be when $\log L \simeq$~-2.0).
{\it Top right panel}: Same as the top left panel, but concentrated in the RCB box area and for the H--R diagram evolution of the CO/He-WD post--merger objects corresponding to models A1 through A4. 
{\it Bottom panels}: Same as the top right panel for the H--R diagram evolution of the CO/He--WD post--merger objects corresponding to models A5 through A8 ({\it bottom left panel}) and B through F ({\it bottom right panel}).}
\label{Fig:HREvols}
\end{center}
\end{figure*}

\begin{figure*}
\begin{center}
\includegraphics[angle=0,width=6in]{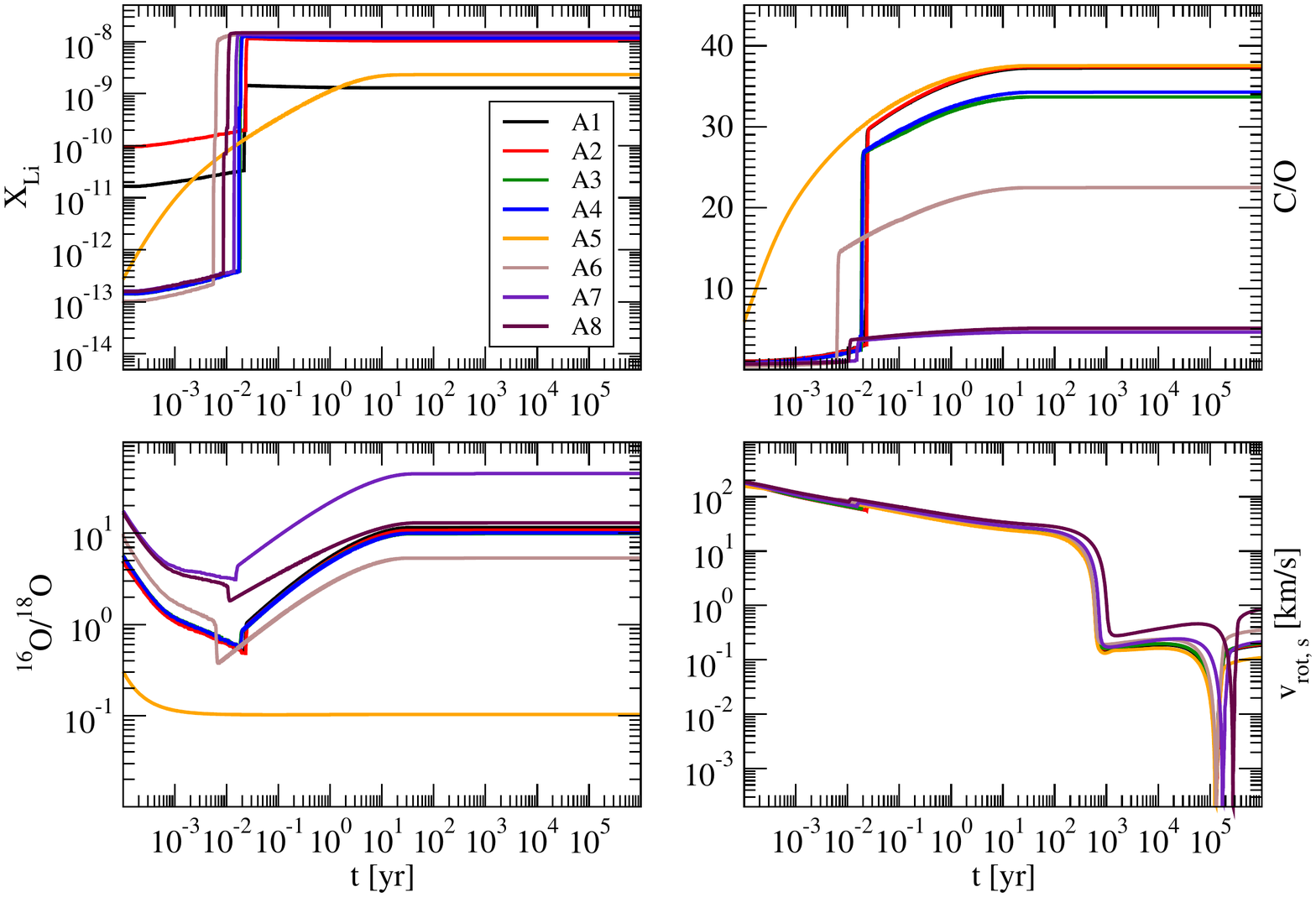}
\includegraphics[angle=0,width=6in]{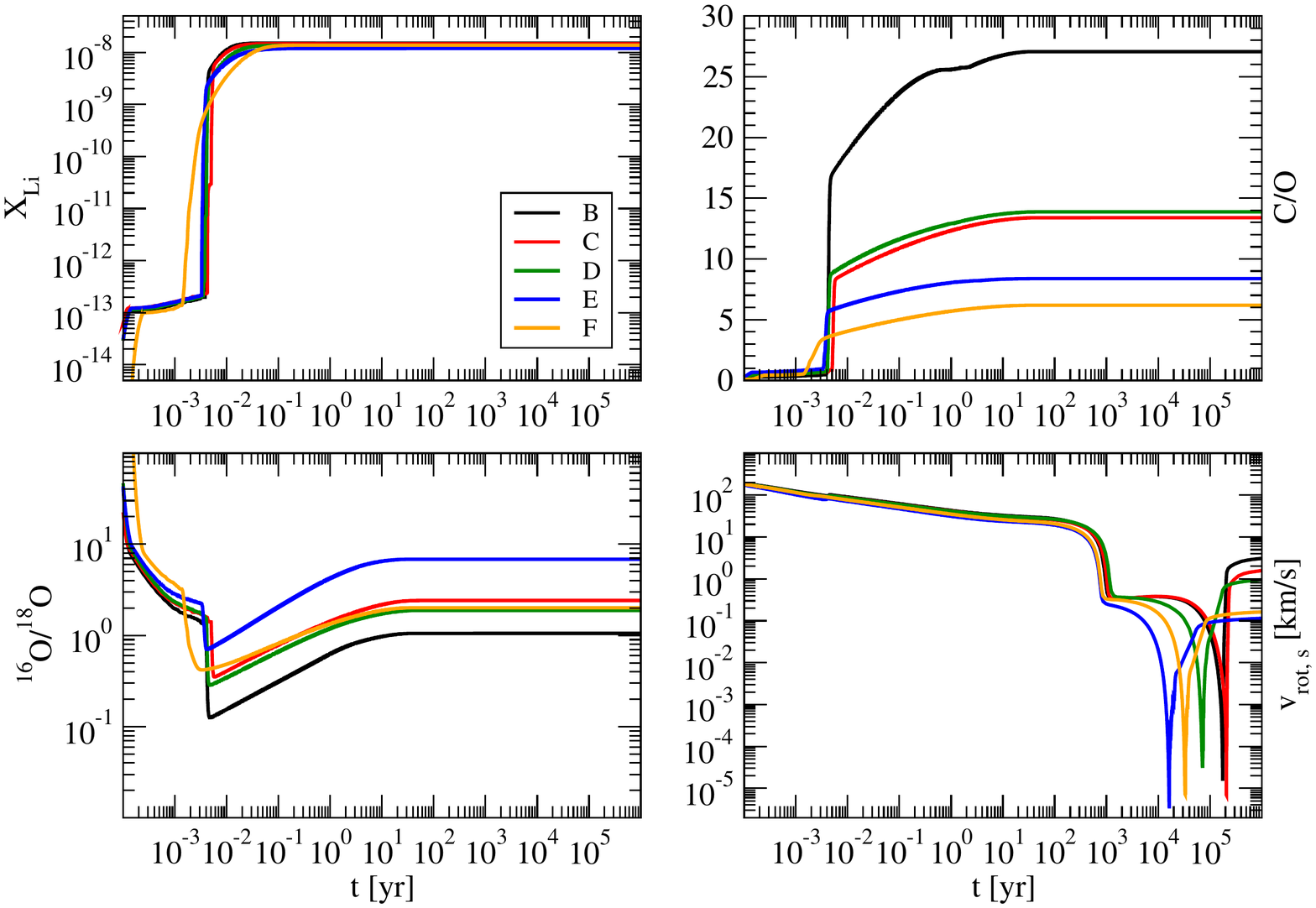}
\caption{Evolution of key RCB parameters for the post--merger CO/He--WD models studied here: 
surface $\rm ^{7}Li$ abundance ({\it upper left panel}), surface $\rm C/O$ ratio ({\it upper right panel}, surface $\rm ^{16}O/^{18}O$ ratio
({\it lower left panel}) and surface equatorial rotational velocity $ v_{\rm rot,f}$ ({\it lower right panel}). }
\label{Fig:ParamEvols}
\end{center}
\end{figure*}

\begin{figure*}
\begin{center}
\includegraphics[angle=0,width=5.5in,clip]{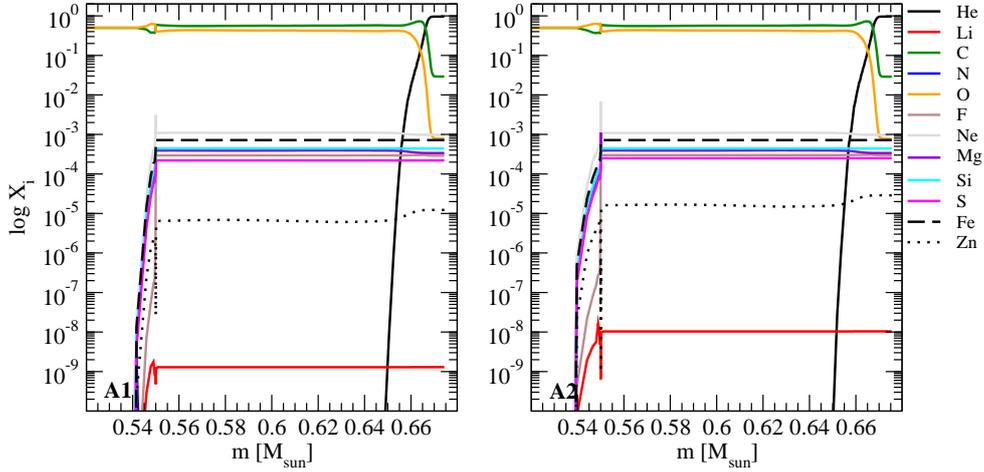}
\caption{Abundance profiles of the final WD remnants at the late--stage in the evolution of the CO/He post--merger objects for Models A1 and A2  (point 4 in the H--R diagram plots).}
\label{Fig:FinalAbund}
\end{center}
\end{figure*}

\setcounter{table}{3}
\begin{table*}
\begin{flushleft}
\caption{Basic properties of the new CO/He-WD post--merger RCB models.}
\begin{tabular}{lccccccccccccc}
\hline
Model&
$M_{\rm i}$&
$M_{\rm CO}$&
$\rm log(\frac{R_{\rm i}}{R_{\odot}}$) &
$\frac{v_{\rm rot,i}}{v_{\rm rot,c}}$ &
$\rm log(X_{\rm H})$ &
$t_{\rm 1}$ &
$t_{\rm 2}$ &
$t_{\rm 3}$ &
$t_{\rm 4}$ &
$t_{\rm 5}$ &
$\rm M_{\rm RCB}$ &
$v_{\rm rot}$\\
&($\rm M_{\odot}$)&($\rm M_{\odot}$)&&&&(10$^5$yr)&10$^5$(yr)&(10$^5$yr)&(10$^5$yr)&(Gyr)&($\rm M_{\odot}$)&(km~s$^{-1}$)
\\
\hline
A1 &0.80&0.55&-1.12&0.2&-99 &0.009&1.256&1.366&1.711&0.18&0.80& 0.15  \\
A2 &0.80&0.55&-1.12&0.2&-9 &0.009&1.284&1.395&1.746&0.19&0.80& 0.16  \\
A3 &0.80&0.55&-1.10&0.2&-5 &0.009&1.270&1.382&1.755&0.16&0.80& 0.16  \\
A4 &0.80&0.55&-1.10&0.0&-5 &0.009&1.265&1.377&1.760&0.19&0.80& 0.00  \\
A5 &0.80&0.55&-1.22&0.2&-5&0.009&1.235&1.339&1.692&0.16&0.80& 0.13  \\
A6 &0.80&0.55&-0.99&0.2&-5   &0.009&1.282&1.400&1.795&0.19&0.80& 0.19  \\
A7 &0.80&0.53&-1.09&0.2&-5   &0.012&1.720&1.846&2.260&0.16&0.80& 0.17  \\
A8 &0.80&0.50&-1.05&0.2&-5   &0.016&2.759&2.922&3.374&0.17&0.80& 0.28  \\
\hline
B  &0.85&0.55&-0.94&0.2&-5    &0.015&1.706&1.845&2.047&0.12&0.85& 0.32 \\
C  &0.90&0.55&-0.90&0.2&-5    &0.015&2.038&2.137&2.304&0.14&0.90& 0.32 \\
D  &0.95&0.55&-0.94&0.2&-5    &0.016&0.742&0.862&1.637&0.16&0.95& 0.37  \\
E  &1.00&0.60&-0.80&0.2&-5   &0.013&0.163&0.187&0.484&0.16&0.99& 0.24  \\
F  &1.05&0.60&-0.74&0.2&-5   &0.011&0.331&0.371&0.743&0.15&1.05& 0.33  \\
\hline
\end{tabular}

The ``large'' network {\tt mesa\_75.net} network was used for all the Models except A5 for which the ``reduced'' choice corresponding to the {\tt sagb\_NeNa\_MgAl.net} was used. See \S~\ref{StellEng} for more details.
The timescales (the time in years since the merger event) correspond accordingly to:  $t_{\rm 1}$, the time to reach the RCB phase (taken to be when the model reaches its maximum luminosity), $t_{\rm 2}$, the time
it takes the model to exit the RCB box (when $\log T_{\rm eff} =$~3.9), $t_{\rm 3}$, the time for the model to reach  $\log T_{\rm eff} =$~4.6 covering the hot RCB stars and the eHe (extreme Helium) stars, and
$t_{\rm 4}$, the time to get to the end of the nearly--constant luminosity post--RCB phase (when $\log L$ starts to drop significantly)
and (E) time for the model to become a new WD (taken to be when $\log L \simeq$~-2.0).

\label{T3}
\end{flushleft}
\end{table*}

\setcounter{table}{4}
\begin{table*}
\begin{flushleft}
\caption{Lifetimes and Expected Number of RCB Stars}
\begin{tabular}{lcccccccccccccc}
\hline
Model&$t_{\rm 2}$&$t_{\rm 3}$&$t_{\rm 3}$-$t_{\rm 2}$&$\rm M_{\rm RCB}$ at $t_{\rm 2}$&$\rm M_{\rm RCB}$$^a$&\# RCB$^b$&\# EHe$^b$\\
&(10$^5$yr)&(10$^5$yr)&(10$^5$yr)&($\rm M_{\odot}$)&($\rm M_{\odot}$)&&&
\\
\hline
A1 &1.256&1.366&0.11&0.80&0.68&226&20\\
A2 &1.284&1.395&0.11&0.80&0.68&231&20\\
A3 &1.270&1.382&0.11&0.80&0.68&229&20\\
A4 &1.265&1.377&0.11&0.80&0.68&228&20\\
A5 &1.235&1.339&0.10&0.80&0.68&222&19\\
A6 &1.282&1.400&0.12&0.80&0.68&231&21\\
A7 &1.720&1.846&0.13&0.80&0.80&310&23\\
A8 &2.759&2.922&0.16&0.80&0.70&497&29\\
\hline
B  &1.706&1.845&0.14&0.85&0.64&307&25\\
C  &2.038&2.137&0.10&0.90&0.66&367&18\\
D  &0.742&0.862&0.12&0.95&0.61&134&22\\
E  &0.163&0.187&0.02&0.99&0.61&29&4\\
F  &0.331&0.371&0.04&1.05&0.63&60&7\\
\hline
\end{tabular}

$^a$RCB mass when He--burning ends.\\
$^b$Predicted numbers of RCB and EHe stars in the Galaxy using the lifetimes predicted by the {\it MESA} models. See discussion in Section 5.5.

\label{T4}
\end{flushleft}
\end{table*}

\section{Post--merger evolution}\label{PMEvol}

\subsection{Grid of post--merger models}

We computed the evolution of a suite of CO/He post--merger objects spanning the relevant parameter space in order to investigate the sensitivity of RCB properties to initial conditions and physical mechanisms. In particular, we tracked the evolution of post--merger objects with masses 0.80, 0.85, 0.90, 0.95, 1.00 and 1.05~$M_{\odot}$. For the 0.80~$M_{\odot}$ model, we modeled cases
varying the initial radius, rotation rate, H abundance in the He--WD, choice of nuclear reaction network, and CO-- to He--WD mass ratio. 

The standard initial conditions that we adopted for all of our post--merger models are:\\
$\bullet$ Radius = $-1.12 < \log(R_{\rm i} / R_{\odot}) < -0.74$\\ 
$\bullet$ Peak He--burning shell temperature $3-5 \times 10^{8}$~K\\
$\bullet$ Rotation rate ($v_{\rm rot,i}/v_{\rm rot,c}$) = 20\% of the critical Keplerian value at the equator\\
$\bullet$ He--WD H mass fraction = $2.8 \times 10^{-5}$\\
$\bullet$ Nuclear reaction network ({\tt mesa\_75.net})\\
$\bullet$ Standard prescriptions for radiatively--driven mass--loss \citep{1975psae.book..229R,1995A&A...297..727B}\\

The initial $M_{\rm CO}$ for models with total mass $< $1~$M_{\odot}$ is chosen to be 0.55~$M_{\odot}$ while for models with $> $1~$M_{\odot}$ it is 0.6~$M_{\odot}$. For the 0.80~$M_{\odot}$ model, we varied the initial conditions to include the following variants: a larger initial radius of  $\log(R_{\rm i} / R_{\odot}) =$-0.72 (and thus lower He-burning shell peak temperature), a case with no rotation, and cases with initial He--WD H mass fractions of 0, $10^{-6}$, $10^{-7}$, and $10^{-10}$. A case using the ``reduced'' ({\tt sagb\_NeNa\_MgAl.net}) network was also run to compare the differences between using the large and small networks. In all cases, we used the standard parameterized treatment for angular momentum transport and chemical mixing due to the effects of rotation that are available in {\it MESA} \citep{2000ApJ...528..368H,2005ApJ...626..350H}. 

As seen in our previously published hydrodynamic simulations \citep{staff2012,staff2018}, the post--merger object is decidedly ellipsoidal. \citet{2011ApJ...737L..34L} suggest that this shape remains at later times in the form of an accretion disk. However, the star becomes spherical on the thermal timescale which is roughly the gravitational potential energy divided by the luminosity of the star. This was calculated in our models using the first two timesteps of the {\it MESA} evolution after the ``stellar engineering" phase described above. Thus, the estimated thermal timescale for the star to become spherical is quite short, $\sim$3--15 yr.

\begin{figure*}
\begin{center}
\includegraphics[angle=0,width=7in,clip]{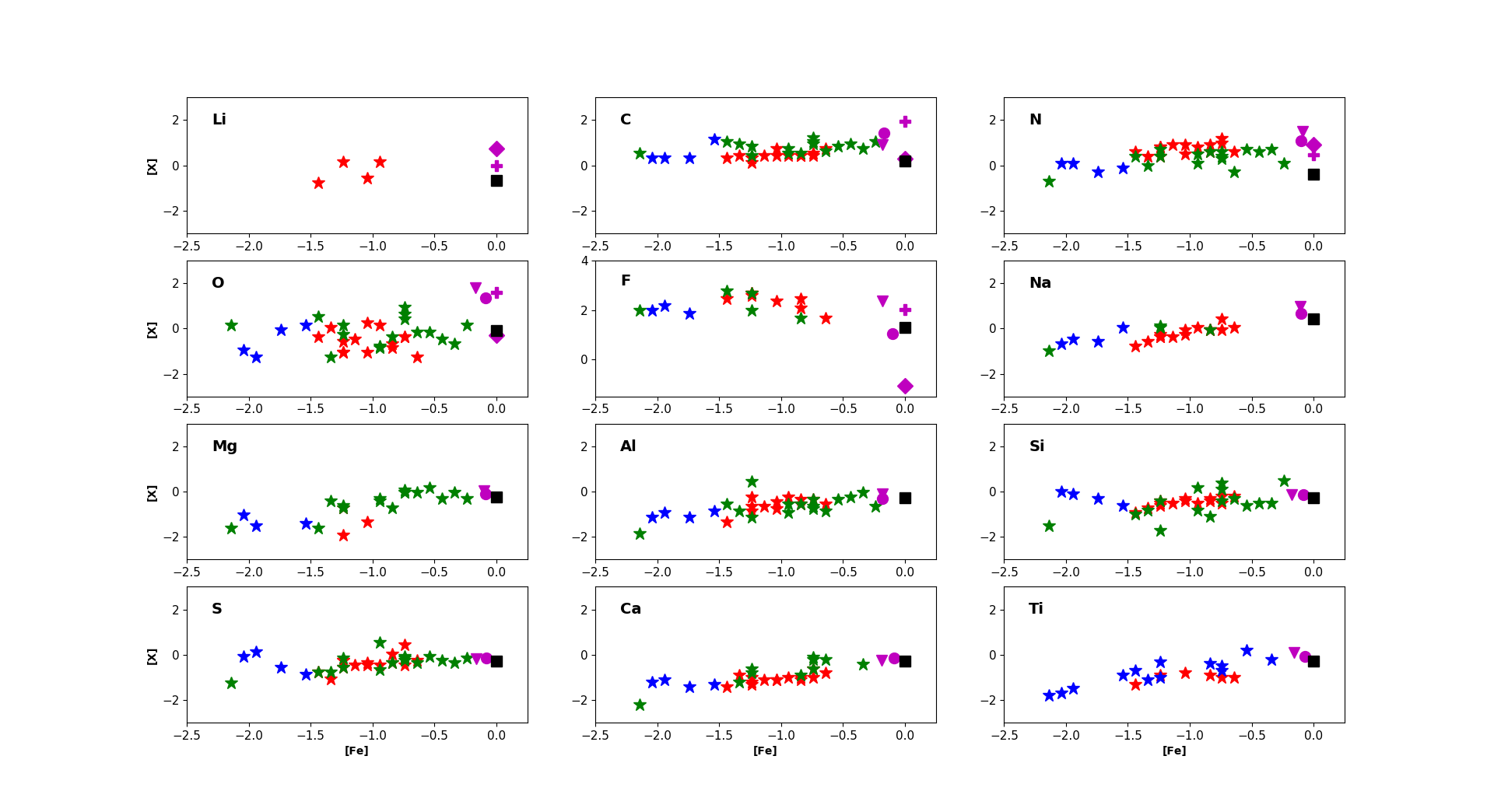}
\caption{The observed surface abundances of RCB and EHe stars
\citep[][and references therein]{2011MNRAS.414.3599J} compared
with the results of our MESA model A1. [X] and [Fe] have the usual definitions representing the logarithmic abundances by number relative to solar \citep{Lodders_2003,2011MNRAS.414.3599J}. The symbols represent majority RCB (red stars), minority RCB (blue stars) and EHe (green stars). The black square in each panel is the abundance from model A1.  The magenta symbols correspond to 
results from other sources in the literature: filled triangle and circle symbols correspond to results presented by \citet{2013ApJ...772...59M}, plus symbols to results by \citet{2011ApJ...737L..34L} and diamond symbols to results by \citet{2014MNRAS.445..660Z}.
}
\label{Fig:[fe]}
\end{center}
\end{figure*}

For all of the models, we computed the evolution through the RCB phase and beyond to the point when the post--merger object forms a new WD remnant. Our computations provide estimates of the relevant time--scales (post--merger to RCB phase, time spent as an RCB star, and timescale to evolve to a cool WD), abundances, as well as other physical properties that can be used for direct comparison with observations. Tables~\ref{T3}--\ref{T4} and Figure~\ref{Fig:[fe]} detail the results for all of the models. All abundances quoted in the tables are calculated
by using $\log \epsilon(X) = \log(X) - \log(\mu_{\rm X}) + 12.15$ where $X$ is the surface mass fraction of an element as calculated in {\it MESA} and $\mu_{\rm X}$ is the mean atomic mass of the element \citep{1998A&A...332..651A,Lodders_2003}. Figures~\ref{Fig:HREvols} and \ref{Fig:ParamEvols} illustrate the evolution of the models in the H--R diagram as well as the evolution of key parameters related to the properties of RCB stars. Figure~\ref{Fig:FinalAbund} shows the compositions of the final WD
remnants. 


As seen in Table~\ref{T4}, the He--burning turn--off mass ranges from 0.61 $M_{\odot}$ for the most massive stars up to 0.7 $M_{\odot}$ for the least massive stars. The He--burning persists until the evolution is near point 2 in Figure 3 ($\sim$10$^5$ yr) when the models start to accelerate toward hotter temperatures in the HR diagram.


\subsection{Abundances and Isotopic Ratios}


The RCB stars have a wide--range of unusual characteristics to which the formation scenarios of a merger of a CO/He white dwarf (WD) binary and a final helium--shell flash have been applied. Both scenarios can account for the hydrogen deficiency, stellar absolute brightness, and some of the measured abundances including the excess s--process elements. However, the WD merger can better explain the large masses, long lifetimes, lack of binarity, relatively low C/He ratio, large measured $^{12}$C/$^{13}$C ratio, and greatly enhanced $^{18}$O and $^{19}$F  seen in the RCB stars \citep{Asplund:2000qy,Clayton:2007ve,2012JAVSO..40..539C}. 
In addition to the previously discussed abundance signature of RCB stars, there is a wide range of H abundances in the RCB stars. There is also an anti--correlation between H and Fe abundances in the RCB stars \citep{Asplund:2000qy}. The RCB stars can be roughly divided into a majority group which share similar abundances, and a small minority of stars, which show extreme abundance ratios, particularly Si/Fe and S/Fe \citep{Asplund:2000qy}. The results of previous studies compared to the observed RCB abundances can be seen in  Figure 2 of \citet{2011ApJ...737L..34L}, Figure 12 of \citet{2013ApJ...772...59M}, Figure 14 of \citet{2014MNRAS.445..660Z}, and Figure 5 of \citet{Menon_2018}. These can be compared to our results in Figure~\ref{Fig:[fe]}. The abundances are plotted against [Fe]. The measured values of [Fe/H] in 18 RCB stars range from 5.5 to 6.9 \citep{Asplund:2000qy}. The Solar value is 7.5 \citep{Lodders_2003}. The low Fe abundances seen in RCB stars has been ascribed to the progenitor stars being slightly metal poor, to dust condensation and separation from the gas, and to additional nucleosynthesis. The high measured Si/Fe and S/Fe ratios seem to require the latter. But none of these suggestions can explain the Fe abundance \citep{Asplund:2000qy}. The Fe abundances modeled in all of the RCB star studies is the same as the initial abundances because Fe does not take part in the nucleosynthesis at these temperatures. 

\subsubsection{Nitrogen}
\begin{figure*}
\begin{center}
\includegraphics[angle=0,width=4in,clip]{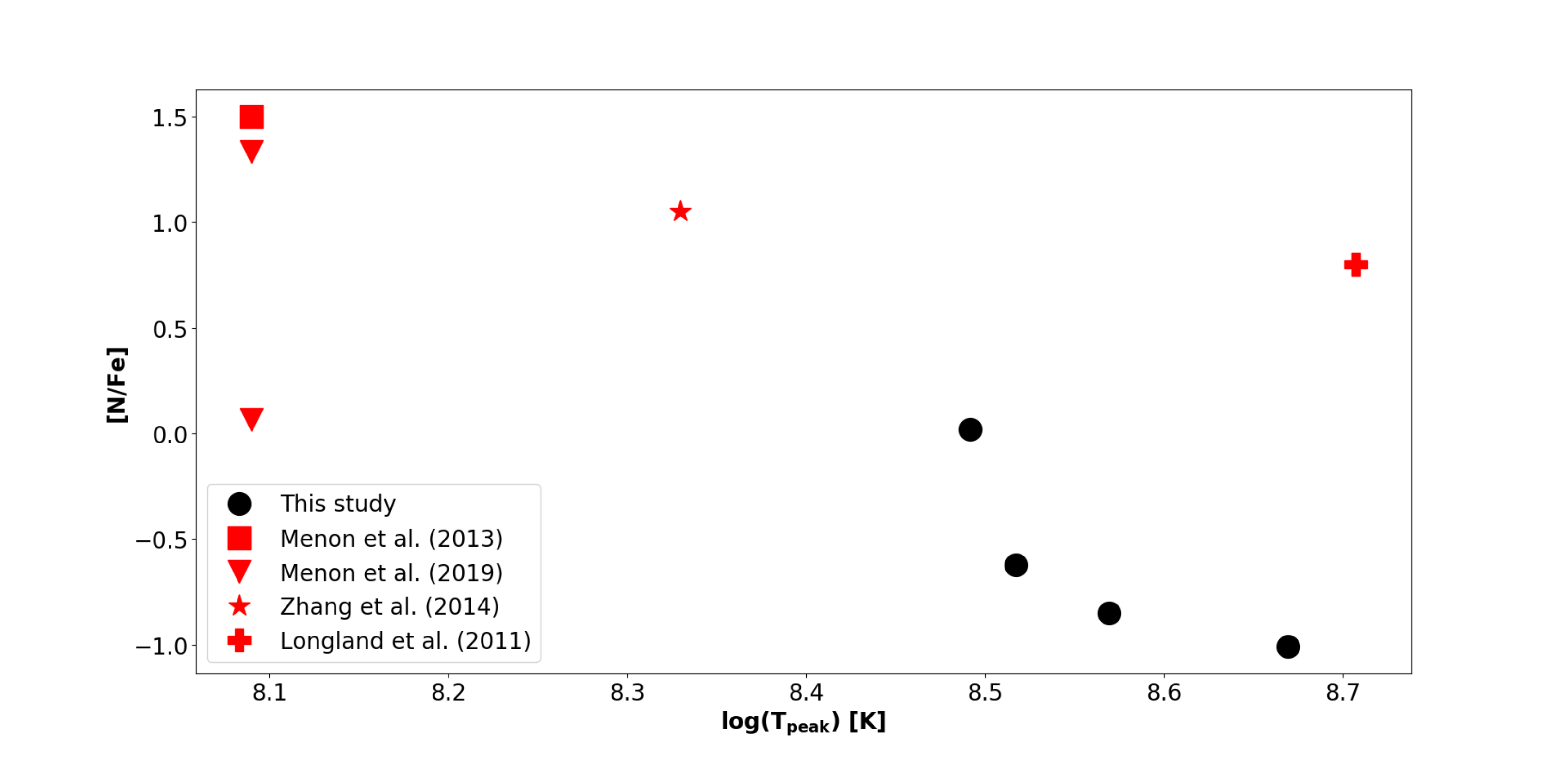}
\caption{The dependence of $^{14}$N surface abundance plotted against peak temperature in the He-burning shell for our models as well as other studies.
}
\label{Fig:nitrogen}
\end{center}
\end{figure*}

To investigate the sensitivity of the resulting RCB model surface abundances of N on the temperature of the He--burning shell, we ran a set of models that are variants of model A5, using the reduced network for speed and efficiency. In particular, by controlling the amount of entropy injected in the He envelope during the MESA engineering step, we were able to control the size of the post-merger object and, as a result, the maximum temperature in the He-burning shell. We find that
the abundances of the CNO elements and N, in particular, are sensitive to the temperature in the He-burning shell. Due to the low H fraction characteristic of these stars, there is no possibility of CNO cycle replenishment. Figure~\ref{Fig:nitrogen} shows the surface $^{14}$N abundance plotted against the peak He-burning shell temperature for our models as compared with other studies \citep{2013ApJ...772...59M,Menon_2018,2014MNRAS.445..660Z}. In general, lower temperature leads to better agreement with the observations but the results of  \citet{2011ApJ...737L..34L} do not follow this trend.
The sensitivity of the results highlights the importance of the nuclear reaction network choice as well as the actual density/temperature structure of the post-merger object. MESA stellar engineering yields a temperature profile that is steeper moving outwards from the peak temperature in the He-burning shell than that suggested by angle-averaged profiles obtained from 3D merger simulations. We aim to investigate the effects of more realistic post-merger structure and more complete nuclear reaction networks on surface RCB abundances in more detail in a future paper.

\subsubsection{$^{18}$O}
The isotope, $^{18}$O, can be enhanced during partial He--burning due to the
$^{14}N$($\alpha,\gamma$)$^{18}$F($\beta^+$)$^{18}$O reaction chain if it is not destroyed by $^{18}$O($\alpha,\gamma$)$^{22}$Ne \citep{Warner:1967lr,1986hdsr.proc..127L,Clayton:2007ve}. 
The models produced by \citet{2011ApJ...737L..34L} show enhanced $^{18}$O and $^{19}$F but not as high as observed in some HdC and RCB stars. In \citet{2014MNRAS.445..660Z}, the models show similar values of $^{18}$O/$^{16}$O to observations, but do not produce enough $^{19}$F to enhance its abundance. 
\citet{2013ApJ...772...59M,Menon_2018} produced models that show enhanced $^{18}$O and $^{19}$F, although the latter is less enhanced than the observations suggest. As noted above, the stellar structure used in Menon is less realistic than in the other studies.

 Our new {\it MESA} modeling produces $^{18}$O/$^{16}$O ratios of $\sim$0.1--40. The observed ratios in the RCB stars are as high as 20 (the Solar value is 1/500) \citep{Clayton:2007ve,Garcia-Hernandez:2010fk}. Fluorine is also enhanced by $^{18}$O(p,$\gamma$)$^{19}$F \citep{Pandey:2008eu}. 
Our lower mass models produce, if anything, too much $^{18}$O as the ratios are a bit higher than observed but do a very good job of reproducing the enhancement of $^{19}$F. 
We find that the abundance of $^{18}$O is lower in our higher mass models which will have higher temperatures in the He-burning shell and may be destroying the $^{18}$O before it can be mixed up to the surface.
None of the studies, including this one, deal with the issue of a significant dredge--up of $^{16}$O during the WD merger which can swamp the amount of $^{18}$O produced \citep{staff2012, staff2018}. 

 Finally, the most obvious difference in terms of $^{18}$O/$^{16}$O is seen for the model involving the ``reduced" network (A5) that results in much smaller ratios than the other models. This is due to the fact that the ``reduced" network contains not only fewer species, but also fewer reactions. Sagb\_NeNa\_MgAl.net contains only  $^{16}$O, $^{17}$O, and $^{18}$O and only 14 of the possible 32 reactions that involve them. Over all the ``reduced" network only includes 52 of the total 185 possible reactions linking the species in Table \ref{T2}.
 Sagb\_NeNa\_MgAl.net neglects all neutron reactions except $^{13}$C($\alpha$,n)$^{16}$O.
 The larger, {\it mesa\_75.net} network includes $^{14}$O-$^{18}$O, and over 78 reactions involving various species of oxygen, as well as all reactions linking the species in Table \ref{T1}. In general these reduced networks are useful to study nucleosynthesis only when combined with larger nets in post-processing, however, this still faces the limitations discussed elsewhere in the paper.

\subsubsection{Lithium}

The appearance of Lithium in the atmospheres of some RCB stars has been considered a strong vote for the final--flash scenario. The abundance of Li in the atmosphere of the final--flash star, Sakurai's object, was actually observed to increase with time \citep{Asplund:1999bh}. Simplistically, one would think that any Li present would be destroyed by the temperatures necessary to produce $^{18}$O.
But a few RCB stars, including R~CrB, itself, show significant Li in their atmospheres \citep{Rao:1996oq,Asplund:2000qy,Kipper:2006fk}. 
\citet{Renzini:1990wd} suggested that the ingestion of the H--rich envelope leads to Li--production through the Cameron--Fowler mechanism ($^3He(\alpha,\gamma)^7Be$ then
$^7Be(e^-,\nu)^7Li$) \citep{Cameron:1971lr}.
\citet{Asplund:2000qy} suggest that the absence of Li in most RCB stars is due to inefficient production or that the new Li is destroyed by exposure to high temperatures before it can be mixed up to the surface.

Since Li is present in all of our models, it is not clear why Li is present in only a handful of RCB stars. In our models, the initial $^3$He mass fraction in the CO--WD is of the order of 10$^{-9}$, which is in the agreement with values in \citet{2014MNRAS.445..660Z}. \citet{2012A&A...542A.117L} connect the observed Li abundance with viewing angle given the non--spherical nature of the post--merger object. They suggest that Li forms through the Cameron--Fowler mechanism in the outer parts of the forming RCB star and at later times resides mainly in a thick accretion disk. Then, if seen side--on, the photosphere of the star is hidden behind the thick accretion disk where you get the enhancement of Li, but when observing RCB stars face--on, radiation of the star itself will dominate the observers measurement and hence, Li abundance measurements will be low. In {\it MESA}, since it is an 1D spherically symmetric code, we follow the evolution along the equatorial plane (effectively including the extended corona/disk structure dominated by He--WD material). Li survives in our models because it is convectively transported out to the safer, lower--temperature regions in the corona on time--scales faster than the ``destruction" time--scales. 



The resulting Li abundance is an enlightening commentary on the competing timescales of nucleosynthesis and convective mixing in light of the extreme hydrogen deficiency. Li survives in our models because it is safely transported out to the lower--temperature corona regions before it can be converted to 2$\alpha$ via the Lithium p-p chain which would ordinarily dominate. Due to the fully-coupled physics and nucleosynthesis, the MLT theory used in these {\it MESA} models implies a shorter timescale for the convective mixing than the Li destruction via burning.

\subsubsection{$^{13}$C}
In general, RCB stars have very large values of $^{12}$C/$^{13}$C \citep{2012JAVSO..40..539C}. A He-rich gas at high temperature will burn $^{13}$C($\alpha$,n)$^{16}$O, providing free neutrons for s-processing and producing large values of $^{12}$C/$^{13}$C. The abundance of $^{12}$C in the models is shown in Figure 6.

 In the models of \citet{2014MNRAS.445..660Z}, the $^{13}$C abundance stays almost perfectly steady. This is because the primary $^{13}$C burning reaction $^{13}$C($\alpha$,n)$^{16}O$ is a neutron reaction, but neutrons as a species were absent from their network, and so were any reactions that included them. Neutron reactions are included in our network as described above \citep{Asplund:2000qy}. 
The s-process is predominantly a neutron process,
so it must be included in any network used to model RCB stars. We see that the $^{13}$C burns up rather quickly in our RCB models, and ratios comparable to Zhang's are difficult to achieve. The final abundance of $^{13}$C in all but one of our {\it MESA} models is near zero. 
The other studies get similarly large $^{12}$C/$^{13}$C ratios. 

\subsubsection{The Effects of Rotation}
The comparison between models A3 and A4 is also of interest since it helps illustrate the effect of rotationally-induced mixing on the surface abundances of RCB stars. These two models have identical initial conditions with the exception of rotation: model A4 is non--rotating while model A3 starts evolving with equatorial rotational velocity at 20\% of the critical Keplerian value. The non-rotating model (A4) retains a little more hydrogen on the surface ($X_{\rm H} =$~-16.01 as compared to $X_{\rm H} =$~-16.65 for the rotating model). In addition, the $^{16}$O/$^{18}$O ratio is a little higher for the non--rotating model ($^{12}$C/$^{13}$C = 9.79 versus 9.29 for the rotating model). This is due to the fact that rotation enables enhanced mixing, predominantly due to meridional circulation, that can recycle unburnt hydrogen--rich material from the outer, cooler layers down to the He--burning shell and gradually exhaust the hydrogen. Overall, however, the effect of rotation is found to be minimal on the surface abundances of RCB stars in our models.


\subsection{Lifetimes and the Expected Number of RCB Stars}

It has been suggested that there is an evolution from He/CO-WD binary mergers $\rightarrow$ RCB stars $\rightarrow$ Extreme Helium (EHe) stars $\rightarrow$ Helium-rich, subdwarf O (He--sdO) stars $\rightarrow$ Helium-rich O (O(He)) stars $\rightarrow$ high-mass CO-WD \citep{2008ASPC..391....3J,2008ASPC..391...53J}.
The total number of RCB stars in the Galaxy is still uncertain. 
Recent concerted efforts to find and identify all of the RCB stars in the Galaxy have increased the number known to 120. This number is unlikely to grow by more than a factor of two in the future \citep{2012A&A...539A..51T,2013A&A...551A..77T,2018arXiv180901743T}.
There are only 22 extreme helium (EHe) stars known. \citet{Jeffery_2017} recently reported the identification of the first new eHe star in 40 years. 
There are only 5 low-gravity He--sdO stars and 4 O(He) stars known \citep{2008ASPC..391....3J}.

Population synthesis calculations indicate an RCB birthrate $\sim$10$^{-2}$--10$^{-3}$ yr$^{-1}$ \citep{Han1998,Nelemans_2001,2013A&A...551A..77T,2014MNRAS.445..660Z,Karakas_2015,Yungelson_2016,Brown_2016}.
RCB stars are thought to be $\sim$0.8--0.9 M$_{\sun}$ from pulsation modeling \citep{Saio:2008qe}, and this mass agrees well with the predicted mass of the merger products of a CO- and a He-WD \citep{Han1998}. 
An actual example of a 0.9 M$_{\sun}$ WD, which is probably the product of a binary merger, was recently discovered in a binary system with a G dwarf \citep{Schwab_2016}.

Using a value of 0.0018 yr$^{-1}$ for the birthrate of RCB stars \citep{Karakas_2015}, we can estimate the number of RCB and EHe stars in the Galaxy by multiplying this value times their typical lifetimes.  
Table 4 gives the lifetimes as RCB stars ($t_{\rm 2} \sim10^{5}$ yr) and the lifetimes of the EHe stars ($t_{\rm 3}$-$t_{\rm 2}\sim10^{4}$ yr) as calculated by our {\it MESA} models. 
Predicted numbers of RCB and EHe stars in the Galaxy using the lifetimes predicted by our {\it MESA} models are given in Table 4.
\citet{2014MNRAS.445..660Z} estimates slightly lower lifetimes but assumes a higher birthrate and so gets similar estimates of the number of RCB and EHe stars in the Galaxy. 

There is a very good correspondence between the estimated numbers of RCB and EHe stars in the Galaxy from our {\it MESA} models and the actual numbers. Most of the models estimate 200--300 RCB stars and 20--30 EHe stars. The estimated lifetimes of the RCB and EHe stars are $\sim$10$^5$ and $\sim$10$^4$ yr, respectively. 
These numbers are also consistent with the idea that RCB and EHe stars are related and are the products of He-/CO-WD mergers.

\section{Conclusions}


While our approach is not radically different from previous studies of RCB star evolution, we have combined rotation, mixing, mass-loss for the first time during the pre-RCB and RCB phases, and included
co-processed nucleosynthesis within {\it MESA}. 

The new models presented here do well in the matching the observed abundances in HdC and RCB stars. 
Good examples are the abundance of $^{19}$F and $^{18}$O/$^{16}$O where our predictions are in good agreement with the observations.
Our new models also reproduce the Li abundances seen in some RCB and the $^{12}$C/$^{13}$C ratios seen in most RCB stars. 
We have also discovered that the N abundance depends sensitively on the peak temperature in the He--burning shell. 

The MESA models predict that the ellipsoidal WD merger product becomes spherical in just a few years, and will become an RCB star in less than 1000 yr. The lifetime of an RCB star is 1--2 $\times 10^5$ yr governed by the timescale of He burning. During the RCB phase, the stellar mass is reduced from $\sim$0.8 to $\sim$0.7 M$_{\odot}$. The stars then evolve more quickly as they return to being WDs. They spend $\lesssim10^4$ yr as EHe stars. When combined with recent population synthesis studies which estimated the merger rate for CO/He WD binaries, these lifetimes predict numbers of RCB and EHe stars which are consistent with the number known in the Galaxy.

In future papers, we plan to investigate the effects of varying the metallicity of the progenitor stars, and of increasing the network to include relevant heavier isotopes.

If the RCB stars can definitively be shown to be the products of WD mergers then the study of how the RCB stars evolve will lead to a better understanding of other important types of stellar merger events such as Type Ia SNe.

\section*{Acknowledgements}
We thank the referee for suggestions that improved this paper. We wish to acknowledge the support from the National Science Foundation through CREATIV grant AST--1240655. EC would like to acknowledge the LSU Department of Physics \& Astronomy and College of Science for startup funding and support. ACL was supported by US Department of Energy, Office of Science, Award DE--SC0014231.

\bibliographystyle{mnras}
\bibliography{everything2}







\bsp	
\label{lastpage}
\end{document}